\documentclass[3p,times]{elsarticle}
\usepackage{lineno,hyperref,multirow,graphicx,booktabs,float,amsmath}

\modulolinenumbers[5]
\journal{arXiv}

\begin{document}
\begin{frontmatter}

\title{Quantifying China's Regional Economic Complexity}

\author[inst1,inst2]{Jian Gao\corref{cor1}}
\cortext[cor1]{\emph{E-mail addresses}: gaojian08@hotmail.com (J. Gao)}
\author[inst1,inst2]{Tao Zhou\corref{cor2}}
\cortext[cor2]{\emph{E-mail addresses}: zhutou@ustc.edu (T. Zhou)}

\address[inst1]{CompleX Lab, Web Sciences Center, University of Electronic Science and Technology of China, Chengdu 611731, People's Republic of China}
\address[inst2]{Big Data Research Center, University of Electronic Science and Technology of China, Chengdu 611731, People's Republic of China}

\begin{abstract}
China has experienced an outstanding economic expansion during the past decades, however, literature on non-monetary metrics that reveal the status of China's regional economic development are still lacking. In this paper, we fill this gap by quantifying the economic complexity of China's provinces through analyzing 25 years' firm data. First, we estimate the regional economic complexity index (ECI), and show that the overall time evolution of provinces' ECI is relatively stable and slow. Then, after linking ECI to the economic development and the income inequality, we find that the explanatory power of ECI is positive for the former but negative for the latter. Next, we compare different measures of economic diversity and explore their relationships with monetary macroeconomic indicators. Results show that the ECI index and the non-linear iteration based Fitness index are comparative, and they both have stronger explanatory power than other benchmark measures. Further multivariate regressions suggest the robustness of our results after controlling other socioeconomic factors. Our work moves forward a step towards better understanding China's regional economic development and non-monetary macroeconomic indicators.
\end{abstract}


\end{frontmatter}
\linenumbers
\nolinenumbers

\section{Introduction}

Understanding how economies develop to prosperity and figuring out the best indicators that reveal the status of economic development are long-standing challenges in economics \cite{Eagle2010,Gao2016}, which have far-reaching implications to practical applications. Traditional macro-economic indicators, like Gross Domestic Product (GDP), are widely applied to reveal the status of economic development, however, calculating these economic census-based indicators are usually costly, resources consuming and following a long time delay \cite{Liu2016}. Thanks to the data revolution of the past decades \cite{Einav2014}, a branch of economic research has been moving to data-driven approaches within the methodology of natural science, statistical physics and complexity sciences \cite{Hamermesh2013,Einav2014b,Hidalgo2016}, which makes it possible to introduce new metrics that surpass the traditional economic measures in revealing current economic status and predicting future economic growth, with applications to economic development \cite{Hidalgo2008,Gao2017a}, trading behavior \cite{Preis2013}, poverty \cite{Blumenstock2015,Jean2016}, inequality \cite{Salesses2013,Hartmann2017}, unemployment \cite{Llorente2015,Yuan2016}, and industrial structure \cite{Gao2017a,Hidalgo2007}. Economists and physicists have also introduced a variety of non-monetary metrics to quantitatively assess the country's economic diversity and competitiveness by measuring intangible assets of the economic system \cite{Hidalgo2009,Hausmann2014}, allowing for quantifying the economies' hidden potential for future development \cite{Cristelli2013,Cristelli2015} in near real-time and at low cost.

In recent decades, many works on quantifying the complexity of socioeconomic systems and financial markets have been done by physicists, who have helped to move research in economy forward by introducing physics-related approaches and models into economic and financial studies \cite{Plerou2003,Preis2006}. In particular, as an interdisciplinary field, the econophysics \cite{Stanley1996,Mantegna1999} applies theories and methods that originally developed by physicists to solve problems in economics and statistical finance \cite{Bouchaud2002}. Recently, econophysicists have proposed network measures to reveal the true risks associated with institutions to make financial markets more stable \cite{Battiston2012} and studied the complex correlations and trend switchings in financial time series \cite{Preis2011}. Moreover, economists and physicists have applied network and statistical methods to reshape the understanding of international trade that the knowledge about exporting to a destination diffuses among related products and geographic neighbors \cite{Jun2017}. Besides, some physical processes like iterative refinement and resource-allocation have been widely applied to evaluate online reputation in socioeconomic systems \cite{Gao2015,Gao2017} and to build better recommender systems in e-commerce \cite{Schafer1999,Chen2017}. More recent works on econophysics and complexity are summarized by review papers \cite{Plerou2000,Chakraborti2011,Huang2015} and books \cite{Chakrabarti2007,Sinha2010}.

Towards quantifying the complexity of a country's economy, the pioneering attempt was made by Hidalgo and Hausmann \cite{Hidalgo2009}, who modeled the international trade flows as ``Country-Product'' networks and derived the Economic Complexity Index (ECI) by characterizing the network structure through a set of linear iterative equations, coupling the diversity of a country (the number of products exported by that country) and the ubiquity of a product (the number of countries exporting that product). The intuition behind this new branch of studies is that the cross-country income differences can be explained by differences in economic complexity, which is measured by the diversity of a country's ``capabilities'' \cite{Hidalgo2009,Hausmann2014}. Soon after, Tacchella et al. \cite{Tacchella2012} developed a new statistical approach which defines a country's Fitness and a product's complexity by the fixed points of a set of non-linear iterative equations \cite{Caldarelli2012}, where the complexity of products is bounded by the fitness of the less competitive countries exporting them. Further, Cristelli et al. \cite{Cristelli2015} studied the heterogeneous dynamics of economic complexity and found, in the fitness-income plane, strong explanatory power of economic development in the laminar regime and weak explanatory power in the chaotic regime. Based on this observation, they argued that regressions are inappropriate in dealing with this heterogeneous scenario of economic development and further proposed a selective predictability scheme to predict the evolution of countries. Nevertheless, these economic complexity indicators are not perfect, for example, ECI suffers from criticisms on its self-consistent, Fitness depends on the dimension of the phase space of the heterogeneous dynamics of economic complexity \cite{Cristelli2013,Cristelli2015}, and a new variant of Fitness method, called minimal extremal metric, can perform even better if for a noise-free dataset \cite{Wu2016}. Recently, Mariani et al. \cite{Mariani2015} quantitatively compared the ability of ECI and Fitness in ranking countries and products, and further investigated a generalization of the ``Fitness-Complexity'' metric.

Even though there is a body of literature on inferring complexity \cite{Hausmann2011,Felipe2012} using cross-country data recording world trade flows \cite{Zaccaria2016,Zhu2017}, studies on China's regional economic complexity using firm level data are still missing. On the one hand, previous studies mainly focus on measuring international level economic competitiveness while the regional level complexity within a country is always ignored. In other words, whether the economic complexity can be successfully extended and tested across different scales is still unknown. One the other hand, most of previous economic complexity analysis are based on the world trade data \cite{Hidalgo2009,Tacchella2012}, meaning that industries without exporting products are excluded, such as services. However, not only goods but also services are important to measure economic complexity as the growth in service and its sophistication can provide an additional route for economic growth \cite{Stojkoski2016}. Moreover, China has experienced a great economic expansion during the past decades. However, some questions regarding China's development are still puzzling, for example, how did China grow \cite{Song2011}, what happened to regional development within China \cite{Gao2017a,Goodman2013}, which metric to use in measuring regional economic complexity \cite{Mariani2015}, and what is the predictive power of complexity to regional development and inequality \cite{Hartmann2017,Xie2014}. Fortunately, the development statistical methods (for example, methodology contributed by econophysicists) and the availability of China's firm level data (dataset that includes all types of industries) provide us a promising way to explore the regional economic complexity within a country, and offers us a chance to explore how the non-monetary economic complexity correlates with traditional monetary macroeconomic indicators at the regional level.

In this paper, we study China's regional economic complexity by analyzing publicly listed firm data from 1990 to 2015. We start by estimating the Economic Complexity Index (ECI) of China's provinces based on the structure of the ``Province-Industry'' network. We show that diversified provinces tend to have industries of less ubiquity, and the overall time evolution of the provinces' rankings by ECI is relatively stable and slow, with provinces located along the coast having higher economic complexity. Then, after linking complexity with the economic development and income inequality, we find that ECI is a positive and significant indicator of economic development with higher explanatory power for provinces of lower level of GDP per capita (GDP pc) that located in laminar regime of ECI-ln(GDP pc) plane compared to provinces of high level of GDP pc that located in chaotic regime. Together, ECI finds a negative and significant explanatory power for regional income inequality of China. Moreover, we compare different measures of economic diversity and explore their relationships with monetary macroeconomic indicators. Results suggest that Fitness is comparative with ECI, and they both perform better than Diversity and Entropy in correlating GDP pc. Further, we show the predictive powers of ECI and Fitness are robust by using multivariate regressions after controlling other socioeconomic factors. Our work contributes to the literature of regional economic complexity.

The paper is organised in the following way. Section~\ref{sec2} introduces the data and the implementation of economic diversity metrics. Section~\ref{sec3} presents the results of China's regional economic complexity and its connections with income inequality. Finally, Section~\ref{sec4} provides conclusions and discussion.

\section{Data and Methods}
\label{sec2}

We study the regional economic complexity by using China's publicly listed firm data, which were extracted from the RESSET Financial Research Database, provided by Beijing Gildata RESSET Data Tech Co., Ltd. (http://www.resset.cn). The data set provides basic registration information and financial information of publicly listed firms in two major stock markets (Shanghai and Shenzhen) of China between 1990 and 2015, such as listing date, delisting date, registered address, and industry category. In our data set, there are in total 2690 firms with the registered addresses covering 31 provinces (or municipalities) in China. For the industry, all the firms belong to 70 categories, which correspond to the industry classification issued by the China Securities Regulatory Commission (http://www.csrc.gov.cn) in 2011.

Some macroeconomic indicators of China at the province-level in the period 2000-2015 are also used, including GDP pc, relative income, income inequality, population, urbanization, schooling, innovation and trade. The GDP pc in the national currency (CNY) is used as a monetary metric in measuring the level of economic development. Together, we also use the relative income in urban area (RICU) and in rural area (RICR) with purchasing power parity (PPP) adjustment in China's provinces for the year 2010, which were original reported by Xie and Zhou \cite{Xie2014}. For the income inequality, we use the relative income differences (RICD) as an estimation, which is defined by the ratio of RICU to RICR to measure the level of income inequality between urban and rural areas in China. For the population, we use the resident population at year-end. For the urbanization metric, we use the share of urban area in a province as an estimation. For the metrics of schooling and innovation, we use the ratio of students in higher education in a province and the number of domestic granted patents, respectively. For the foreign trade, we use the total value of imports and exports of destinations and catchments. All these macroeconomic data except for income inequality were extracted from China Statistical Yearbook, published by the National Bureau of Statistics of China (http://www.stats.gov.cn).

\begin{figure*}[t]
  \centering
  \includegraphics[width=0.8\textwidth]{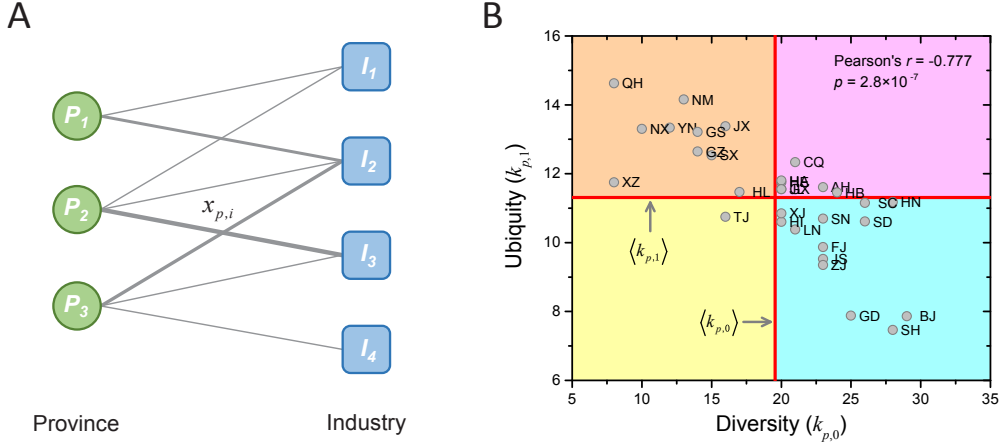}
  \caption{Quantifying regional economic complexity. (\textbf{A}) Illustration of a ``Province-Industry'' bipartite network. The weight of link $x_{p,i}$ is the number of firms in province $p$ that operate in industry $i$. (\textbf{B}) The ``Diversity-Ubiquity'' diagram divided into four quadrants defined by the averaging diversity $\langle k_{p,0}\rangle$ and ubiquity $\langle k_{p,1}\rangle$, as shown by the vertical and horizontal lines, respectively. The abbreviations of province names correspond to Table~\ref{Tab:A1} in Appendix.}
  \label{Fig1}
\end{figure*}

The Economic Complexity Index (ECI) \cite{Hidalgo2009} is proposed based on the intuition that, in our case, sophisticated economies are of high diversity (the number of industries one province has) and have industries with low ubiquity (the number of provinces with that industry), because only a few provinces can diversify into sophisticated industries \cite{Hartmann2017,Hausmann2014,Hausmann2011}. So combining information on the diversity of a province and the ubiquity of its industries will provide a promising way to measure the sophistication of a province's productive structure. To calculate ECI, we first build a ``Province-Industry'' network where the weight of a link is defined as the number of firms belonging to the corresponding industry type and in the corresponding province (see Figure~\ref{Fig1}A for an illustration). Then, we transform this bipartite network into an adjacency matrix $M_{p,i}$, where $M_{p,i}=1$ if province $p$ has the revealed comparative advantage (RCA) \cite{Balassa1965} in industry $i$ ($\text{RCA}_{p,i} \ge 1$) and 0 otherwise \cite{Hidalgo2009,Bahar2014}. Here, we define the RCA as the ratio between the observed number of firms operating in an industry in a province and the expected number of firms of that industry in that province. Formally, the RCA for province $p$ in industry $i$ is defined as
\begin{equation}
    \text{RCA}_{p,i} = \left.{\frac{x_{p,i}}{\sum_{i'}x_{p,i'}}}\middle/ \frac{\sum_{p'}x_{p',i}}{\sum_{i'}\sum_{p'}x_{p',i'}}\right.
    ,
\end{equation}
where $x_{p,i}$ is the number of firms in province $p$ that operate in industry $i$. Further, the diversity of a province $p$ is defined as the number of industries in which the province has the comparative advantage,
\begin{equation}
    \text{Diversity} = k_{p,0} = \sum_{i}M_{p,i}.
    \label{Eq:Div}
\end{equation}
The ubiquity of industry $i$ is defined as the number of provinces with the comparative advantage in that industry,
\begin{equation}
    \text{Ubiquity} = k_{i,0} = \sum_{p}M_{p,i}.
\end{equation}
Finally, the Economic Complexity Index (ECI) of province $p$ is defined as
\begin{equation}
    \text{ECI}_p = \frac{K_p-\langle \vec{K}\rangle}{std(\vec{K})}=\frac{m^{2}K_{p}-m\sum_{p}{K_p}}{\sqrt{m\sum_{p}{(mK_p-\sum_{p}{K_p}})^2}},
    \label{Eq:ECI}
\end{equation}
where $m$ is the number of provinces, $\langle \cdot \rangle$ and $std(\cdot)$ are respectively functions of mean value and stand deviation that operate on the elements in vector $\vec{K}$, and $\vec{K}$ is the eigenvector associated with the second largest eigenvalue of the matrix \cite{Caldarelli2012}
\begin{equation}
    \tilde{M}_{p,p'} = \frac{1}{k_{p,0}}\sum_i {\frac{M_{p,i}M_{p',i}}{k_{i,0}}}.
\end{equation}
Indeed, the matrix $\tilde{M}_{p,p'}$ is defined in terms of connecting provinces who have similar industries, weighted by the inverse of the ubiquity of an industry ($k_{i,0}$) and normalized by the diversity of a province ($k_{p,0}$) \cite{Hausmann2014}.

The Fitness Index \cite{Tacchella2012} is based on the idea that, i) a diversified province gives limited information on the complexity of industries, and ii) a poorly diversified province is more likely to have a specific industry of a low level sophistication. Therefore, a non-linear iteration is needed to bound the complexity of industries by the fitness of the less competitive provinces having them \cite{Cristelli2013,Cristelli2015}. Here, the Fitness of province is proportional to the number of its industries weighted by their complexity. In turn, the complexity of industry is inversely proportional to the number of provinces who have this industry (similar methods were early proposed to deal with recommender systems \cite{Gao2017,Zhou2011}. The coupling of the province $p$'s Fitness ($F_p$) to the industry $i$'s complexity ($Q_i$) is summarized in the following non-linear iterative scheme:
\begin{equation}
    \left\{
    \begin{aligned}
        \tilde{F}_{p}^{(n)} & = & \sum_{i}{M_{p,i}Q_{i}^{(n-1)}} \\
        \tilde{Q}_{i}^{(n)} & = & \frac{1}{\sum_{p}{M_{p,i}\frac{1}{F_{p}^{(n-1)}}} }
    \end{aligned}
    \right.
    ,
    \label{Eq:Fit}
\end{equation}
where $\tilde{F}_{p}^{(n)}$ and $\tilde{Q}_{i}^{(n)}$ are respectively normalized in each step by $F_{p}^{(n)}=\tilde{F}_{p}^{(n)} / \langle \tilde{F}_{p}^{(n)}\rangle$ and $Q_{i}^{(n)}=\tilde{Q}_{i}^{(n)} / \langle \tilde{Q}_{i}^{(n)}\rangle$ given the initial condition $F_{p}^{(0)}=1$ and $Q_{i}^{(0)}=1$. The non-linear iterations go until the stationary state is reached, and the final Fitness value reflects complexity.

\begin{figure*}[t]
  \centering
  \includegraphics[width=0.8\textwidth]{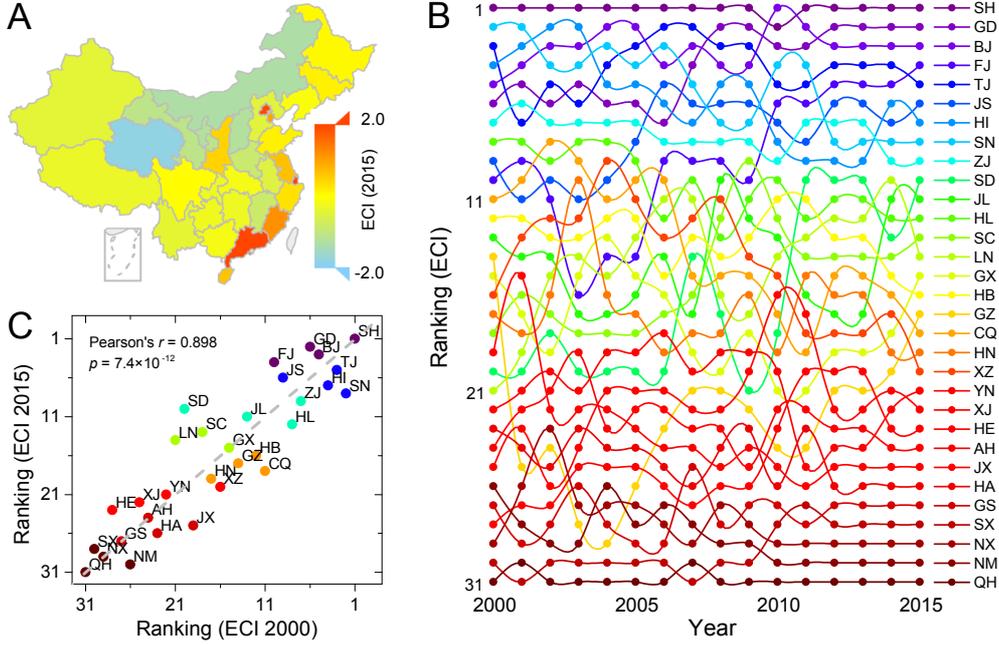}
  \caption{China's regional economic complexity and the evolution of provinces' rankings. (\textbf{A}) Map of China's regional Economic Complexity Index (ECI). The color denotes the value of ECI in 2015. (\textbf{B}) Time evolution of all provinces' rankings by ECI from 2000 to 2015. (\textbf{C}) Relationship between rankings by ECI in 2000 and 2015. The gray dash line is the diagonal line. The abbreviations of province names correspond to Table~\ref{Tab:A1} in Appendix.}
  \label{Fig2}
\end{figure*}

\section{Results}
\label{sec3}

In this section, we first report the China's regional economic complexity ECI and its time evolution. Then, we show the heterogeneity of the levels of economic development in relation to the value of the economic complexity. Next, we provide simple comparisons among different measures of economic diversity, especially ECI and Fitness, and show how they correlate other monetary macroeconomic indicators. Finally, we check the robustness of the predictive power of the economic complexity for the economic growth by using multivariate regressions.

\subsection{Regional Economic Complexity}

The Economic Complexity Index (ECI) measures the regional economic structure by combining province's diversity and industry's ubiquity. To check the intuition behind ECI that sophisticated economies are diverse and having industries of low ubiquity, in Figure~\ref{Fig1}B we present the relationship between a province's diversity ($k_{p,0} = \sum_{i}M_{p,i}$) and the averaging ubiquity of industries in which the province has the comparative advantage ($ k_{p,1} = \sum_{i}{(k_{i,0}M_{p,i})}/\sum_{i}{M_{p,i}}$). We find a strong and significant negative correlation between $k_{p,0}$ and $k_{p,1}$ with Pearson's correlation $r=-0.777$ ($p\text{-value}=2.8 \times 10^{-7}$), supporting the hypothesis that diversified provinces tend to have less ubiquitous industries.

Figure~\ref{Fig2}A presents the values of China's regional Economic Complexity Index (ECI) at province level in 2015. We find that provinces located along the coast trend to have higher economic complexity, follow by provinces that located in Southwest and Northeast of China. Figure~\ref{Fig2}B shows the time evolution of the rankings of all provinces between 2000 and 2015 by ECI. It can be seen that provinces with highest and lowest rankings are more stable during that period, with Shanghai (SH), Guangdong (GD) and Beijing (BJ) ranked to the top, and Qinghai (QH), Inner Mongolia (NM) and Ningxia (NX) ranked to the bottom. For the middle rankings, economies in some provinces become more sophisticated such as Shandong (SD) and Fujian (FJ) while some provinces become less complex such as Shaanxi (SN) and Chongqing (CQ). Figure~\ref{Fig2}C compares all rankings by ECI at the starting year 2000 and the ending year 2015. Provinces with an increased ECI rankings locate above the diagonal while provinces with decreased rankings locate under the diagonal. We find that the ECI rankings in 2015 are highly and significantly correlated with those in 2000 with the Pearson's correlation $r=0.898$ ($p\text{-value}=7.4 \times 10^{-12}$), indicating that the overall time evolution of ECI rankings is relatively stable and slow.

\begin{figure*}[t]
  \centering
  \includegraphics[width=0.8\textwidth]{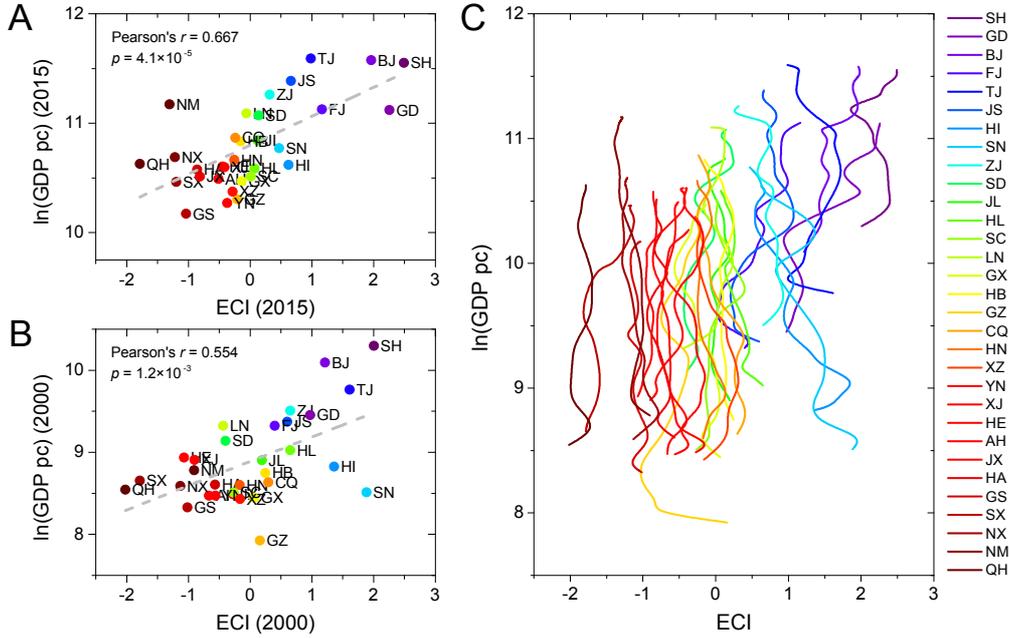}
  \caption{Relationship between economic developmet and economic complexity. (\textbf{A}) and (\textbf{B}) show the positions of provinces in plane of Economic Complex Index (ECI) versus natural logarithm of GDP pc in 2015 and 2000, respectively. The gray dash line is the linear fit of dots. (\textbf{C}) Time evolution of locations of provinces in the ECI-ln(GDP pc) plane from 2000 to 2015. During this period, the better-going and worse-going provinces changed their ECIs, on average, with a value 0.49 and $-0.40$, respectively. The abbreviations of province names correspond to Table~\ref{Tab:A1} in Appendix.}
  \label{Fig3}
\end{figure*}

\subsection{Linking Complexity to Development and Inequality}

The Economic Complexity Index (ECI) is a non-monetary metric which is able to assess the level of development and competitiveness of provinces by measuring intangible assets of economic systems \cite{Hidalgo2009,Cristelli2015,Tacchella2012}. Naturally, we should compare this metric for province intangibles with monetary metric, as the GDP pc, which is traditionally used by economists in measuring the level of economic development. For static observations, Figure~\ref{Fig3}A and \ref{Fig3}B show the locations of provinces in the ECI-ln(GDP pc) plane for 2015 and 2010, respectively. We find that the economic complexity is a positive and significant indicator of economic development, as suggested by the high correlation between ECI and ln(GDP pc) with Pearson's correlation $r=0.667$ ($p\text{-value}=4.1 \times 10^{-5}$) for 2015 and $r=0.554$ ($p\text{-value}=1.2 \times 10^{-3}$) for 2000. Roughly speaking, provinces with larger economic complexity enjoy a higher level of economic development.

To further investigate how economic development depends on the complexity, we move from the static pictures to the dynamics of provinces in the compound ECI-ln(GDP pc) plane from 2000 to 2015. As shown in Figure~\ref{Fig3}C, the dynamics of the provinces in this plane is, to some extent, heterogeneous but with two emergent trends. On the left and central sides, we observe a laminar regime, where ECI is linearly and positively correlated with ln(GDP pc), supporting that ECI is a driving force of economic growth \cite{Cristelli2015}. Countries locating in this laminar regime enjoy a slow but stable economic development. On the right side, we observe a chaotic regime, where the dynamics of provinces are less predictable due to the larger fluctuations of ECI. However, countries locating in this chaotic regime developed much faster and achieved a higher level of economic development. For example, Shaanxi (SN) has a much higher ECI than the other provinces with the same level of GDP pc (for example, QH, SC, JX, AH and YN) in 2000. In the last 15 years, the GDP pc of Shaaxi (SN) increased by a factor of 9.6, leading its ranking by GDP pc jumped from 23 to 14. By comparison, the GDP pc of the other provinces with the same level of GDP pc only grew, on average, by a factor of 7.3. These results suggest that, in the case of China, ECI is a good indicator of future economic development for the provinces with a relatively low level of GDP pc, while the predictive power is reduced for the provinces with a high level of GDP pc. Moreover, we notice that during the considered period the ECIs of some provinces remarkably increased, for example, Shanghai (SH) from 2.01 to 2.49 and Guangdong (GD) from 0.97 to 2.26 while the ECIs of some provinces decreased remarkably, for example, Tianjin (TJ) from 1.61 to 0.99 and Jiangxi (JX) from $-0.27$ to $-0.81$. On average, the ECIs of better-going province (with increasing in ECI) changed 0.49, and the ECIs of worse-going provinces (with decreasing in ECI) changed $-0.40$ from 2000 to 2015. The result suggests that economic complexity and regional development are heterogeneous within a country.

\begin{figure*}[t]
  \centering
  \includegraphics[width=0.8\textwidth]{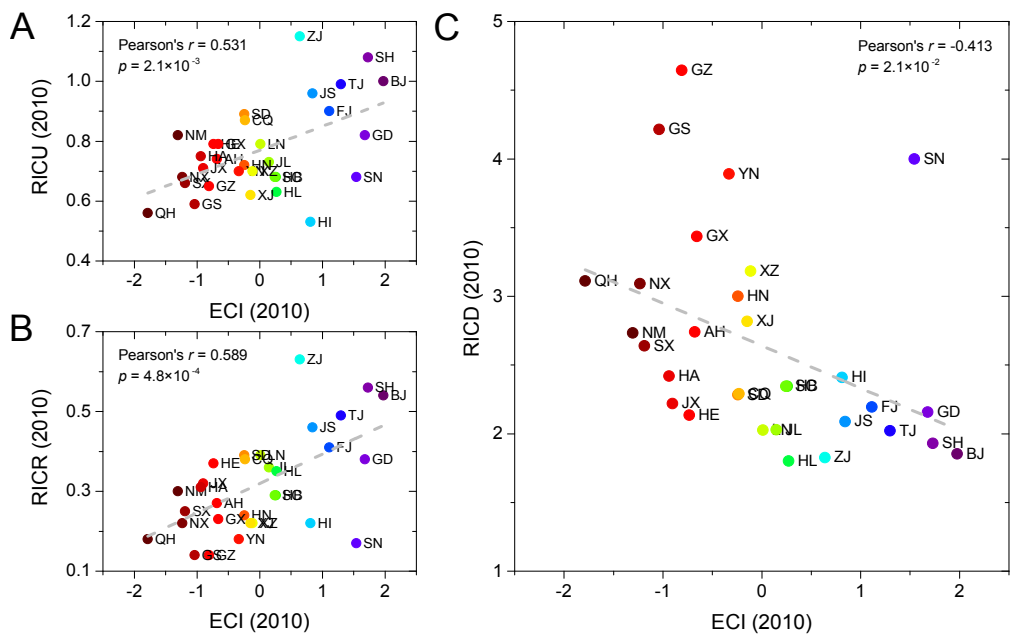}
  \caption{Relationship between economic complexity and income inequality. (\textbf{A}) and (\textbf{B}) are respectively for Economic Complexity Index (ECI) versus relative income at urban area (RICU) and at rural area (RICR) in 2010. (\textbf{C}) Relationship between ECI and relative income differences (RICD). As an estimation of income inequality, RICD is defined by the ratio of RICU to RICR. The gray dash line is the linear fit of dots. The abbreviations of province names correspond to Table~\ref{Tab:A1} in Appendix.}
  \label{Fig4}
\end{figure*}

The income inequality along with economic development has always been a central concern of economists and policy makers in economic theory and policy \cite{Kuznets1955}. With the development of new perspective and economic tools, progresses have been made on explaining income inequality through new data and measures \cite{Hartmann2017,Deininger1996,Li1998}. For example, Hartmann et al. \cite{Hartmann2017} showed that the economic complexity can be a significant and negative indicator of income inequality. We here explore the relationship between economic complexity and income inequality on regional level within China. First, Figure~\ref{Fig4}A and \ref{Fig4}B show how ECI correlates with the relative income at urban area (RICU) and at rural area (RICR), respectively. We find a positive and significant correlation between ECI and the relative income at both urban area (Pearson's correlation $r=0.531$ with $p\text{-value}=2.1 \times 10^{-3}$) and rural area (Pearson's correlation $r=0.589$ with $p\text{-value}=4.8 \times 10^{-4}$). Then, we use the relative income differences (RICD), defined by the ratio of RICU to RICR, as an estimation of income inequality and further show the relationship between ECI and RICD in Figure~\ref{Fig4}C. We find a negative and significant correlation between economic complexity and relative income differences (Pearson's correlation $r=-0.413$ with $p\text{-value}=2.1 \times 10^{-2}$), which is coincided with previous findings based on international trade data \cite{Hartmann2017}. These results suggest that China's economic complexity still has negative explanatory power of regional income inequality, although China's great economic expansion has risen regional disparities significantly higher during the last a few decades \cite{Wan2007,Li2013}. Once again, the results suggest the development of regions in China is not homogenous (see GDP pc in Figure~\ref{Fig3} and income inequality in Figure~\ref{Fig4} for examples), even though the country as whole may experience remarkable increase in economic complexity and development. The observations should cause us to further explore the complexity and development at both national and regional levels.

\subsection{Comparing Different Measures of Economic Diversity}

Thanks to the development of complexity sciences, a variety of metrics have been proposed to measure the diversity of economies regarding their productive structures, including Economic Complexity Index (ECI) \cite{Hidalgo2009}, Fitness Index \cite{Tacchella2012}, Diversity \cite{Hidalgo2009} and Entropy \cite{Shannon1948}. The ECI defines a country's complexity and a product's ubiquity through a set of linear iterative equations. The Fitness Index defines a self-consistent metrics for a province's fitness and a product's complexity through a set of no-linear iterative equations that assess the advantage of diversification. The Diversity is defined by Eq.~(\ref{Eq:Div}), i.e., the number of industries in which one province has the comparative advantage. The Shannon Entropy measures the diversity of industries in which one province has the comparative advantage.

\begin{figure*}[t]
  \centering
  \includegraphics[width=0.8\textwidth]{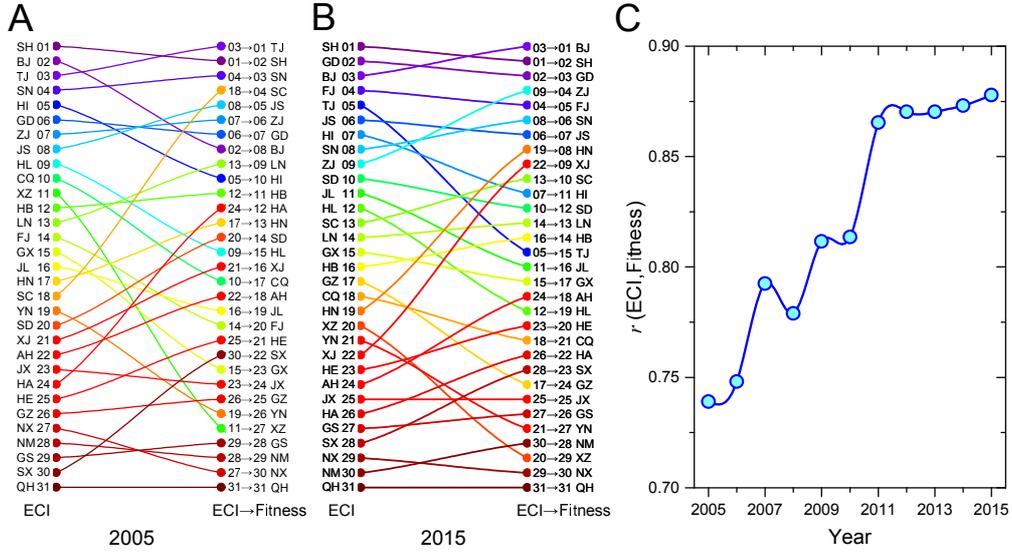}
  \caption{Comparison between Economic Complexity Index (ECI) and Fitness Index. (\textbf{A}) and (\textbf{B}) show the mappings of the rankings of provinces by ECI (left) into Fitness (right) in 2005 and 2015, respectively. (\textbf{C}) The Pearson's correlation coefficient between ECI and Fitness as a function of time. All the positive correlations are significant with $p$-value no more than $10^{-6}$. The abbreviations of province names correspond to Table~\ref{Tab:A1} in Appendix.}
  \label{Fig5}
\end{figure*}

First, we compare the ability of ECI and Fitness on ranking the complexity of China's provinces. Figure~\ref{Fig5}A and \ref{Fig5}B present how the rankings by ECI is mapped into the rankings by Fitness in 2005 and 2015, respectively. In general, we find that ECI and Fitness agree with each other for top rankings and bottom rankings, while the two methods are distinguishable for middle rankings. For example, Hainan (HN) and Xinjiang (XJ) are respectively ranked 19 and 22 by ECI in 2015, while the corresponding rankings are 8 and 9 by Fitness. There are also some provinces that are overestimated by ECI compared to Fitness in 2015, such as Tianjin (TJ: 5$\rightarrow$15), Heilongjiang (HL: 12$\rightarrow$19) and Tibet (XZ: 20$\rightarrow$29). To provide a quantitative comparison between the rankings, in Figure~\ref{Fig5}C we show the Pearson's correlation $r$ between ECI and Fitness as a function of time. We find a positive and significant correlation between the two rankings across all years with $p$-value no more than $10^{-6}$. The correlation is stabilized at about 0.871 since 2011, suggesting that the rankings by ECI and Fitness are, to some extent, consistent and stable. The result is notable since previous studies based on world trade data found inconsistency of ECI and Fitness methods in ranking countries \cite{Cristelli2013,Mariani2015}. Here, our empirical results based on firm data at regional level suggest that the two methods are comparative. Considering that there is no ground truth in rankings in terms of economic complexity and the two methods have distinctive intuitions and formulations, it is hard to identify the best methods in practices, leaving the problem being still complicated. Indeed, the discrepancies of these four measures of economic diversity urge on the development of new regional economic complexity metrics.

\begin{figure*}[t]
  \centering
  \includegraphics[width=0.8\textwidth]{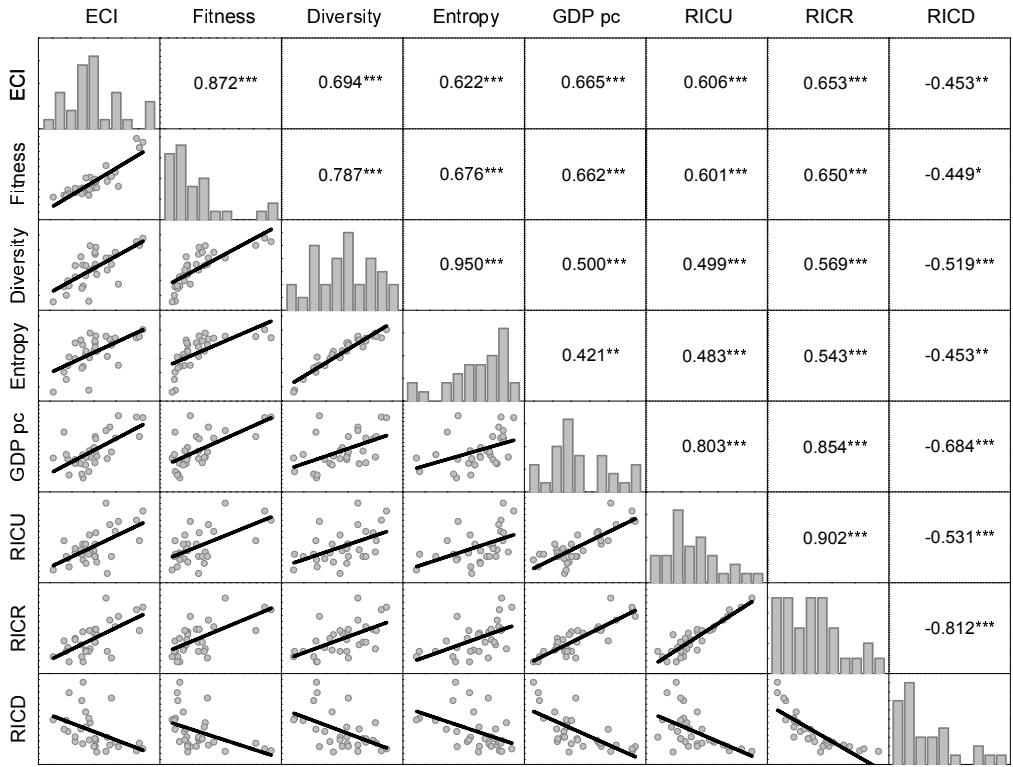}
  \caption{Correlations between different economic diversity measures and economic development as well as income inequality. The economic diversity measures include Economic Complexity Index (ECI), Fitness Index, Diversity and Shannon Entropy. The economic development measures include GDP pc, relative income at urban area (RICU) and relative income at rural area (RICR). The income inequality is estimated by the relative income differences (RICD), defined by the ratio of RICU to RICR. All metrics are averaged over the period 2010-2015 to reduce noises except for RICU, RICR and RICD, which are only for 2010. The matrix diagonal shows the histograms of each variable, the upper triangle shows the Pearson's correlation coefficients between the pair of variables, and the lower triangle shows the corresponding scatter-plots with solid lines representing linear fits. The correlation coefficients are with significant level *$p<0.1$, **$p<0.05$, and ***$p<0.01$.}
  \label{Fig6}
\end{figure*}

Next, we explore the correlations among different measures of economic diversity, economic development and income inequality. As shown in the first four columns of Figure~\ref{Fig6}, all the four economic diversity metrics have positive and significant correlations with each other. Specifically, Fitness is highly correlated with ECI (see the second columns of Figure~\ref{Fig6}), and Diversity is highly correlated with Entropy (see the fourth columns of Figure~\ref{Fig6}). ECI and Fitness have higher explanatory power for GDP pc compared to Diversity and Entropy, as suggested by their larger correlation coefficients, 0.665 for ECI and 0.662 for Fitness (see the fifth column of Figure~\ref{Fig6}). Also, ECI and fitness are better indicators for relative income, compared to Diversity and Entropy (see the sixth and seventh columns of Figure~\ref{Fig6}). Together, we find that the correlation coefficients in RICR column are much larger than the corresponding values in RICU column, suggesting that the relative income in rural area are more explained by economic diversity metrics than in urban area. For the income inequality, we find that all the measures of economic diversity and economic development are negatively and significantly correlated with RICD (see the last column of Figure~\ref{Fig6}), meaning that the more economic diversity and the higher level of economic development, the less income inequality. In particular, RICR has the highest explanatory power for RICD, followed by GDP pc and RICU, suggesting that the relative income in rural area has the potential to be the best negative indicator for income inequality in provinces. The reason why economic complexity measures are less competitive than, for example, RICR, in correlating with RICD is still puzzling, which urges for further exploration. Moreover, we notice that ECI and Fitness are comparable with each other in explaining economic development and income inequality as indicated by their very close correlation coefficients (see the first two rows of Figure~\ref{Fig6}).

\subsection{Robustness Check of Complexity's Predictive Power}

Using bivariate statistics in the above two sections, we have shown the correlations between economic diversity measures and the level of economic development. In this section, based on multivariate regressions, we further explore whether changes in a province's economic diversity are associated with changes in the level of economic development after controlling the effects of other socioeconomic factors like Population, Urbanization, Schooling, Innovation and Trade. If the economic diversity is a good and robust indicator for economic development, we should observe a positive and significant correlation between the former non-monetary metrics (ECI and Fitness) and the later monetary metric (GDP pc).

\begin{table*}[t]
  \centering
  \footnotesize
  \caption{Results of the multivariate regressions for the level of economic development.}
    \begin{tabular*}{\textwidth}{@{\extracolsep{\fill}}lcccccccc}
    \hline
    \multirow{3}[6]{*}{} & \multicolumn{7}{c}{OLS model with dependent variable: ln(GDP pc)}        &  \\
    \cline{2-9}          & \multicolumn{4}{l}{ECI}       & \multicolumn{4}{l}{Fitness} \\
    \cline{2-9}          & (1)   & (2)   & (3)   & (4)   & (5)   & (6)   & (7)   & (8) \\
    \hline
    \multirow{2}[1]{*}{ECI/Fitness} & \multicolumn{1}{l}{0.2736***} & \multicolumn{1}{l}{0.1699***} & \multicolumn{1}{l}{0.1039***} & \multicolumn{1}{l}{0.1398***} & \multicolumn{1}{l}{0.2746***} & \multicolumn{1}{l}{0.1707***} & \multicolumn{1}{l}{0.1270***} & 	\multicolumn{1}{l}{0.1445***} \\
          & \multicolumn{1}{l}{(0.0237)} & \multicolumn{1}{l}{(0.0302)} & \multicolumn{1}{l}{(0.0304)} & \multicolumn{1}{l}{(0.0307)} & \multicolumn{1}{l}{(0.0235)} & \multicolumn{1}{l}{(0.0390)} & \multicolumn{1}{l}{(0.0297)} & \multicolumn{1}{l}{(0.0298)} \\
    \multirow{2}[0]{*}{ln(Population)} &       & \multicolumn{1}{l}{0.0148} &       &       &       & \multicolumn{1}{l}{0.0014} &       &  \\
          &       & \multicolumn{1}{l}{(0.0269)} &       &       &       & \multicolumn{1}{l}{(0.0296)} &       &  \\
    \multirow{2}[0]{*}{Urbanization} &       & \multicolumn{1}{l}{0.7314***} &       &       &       & \multicolumn{1}{l}{0.6349***} &       &  \\
          &       & \multicolumn{1}{l}{(0.1386)} &       &       &       & \multicolumn{1}{l}{(0.1767)} &       &  \\
    \multirow{2}[0]{*}{Schooling} &       &       & \multicolumn{1}{l}{21.488***} &       &       &       & \multicolumn{1}{l}{22.165***} &  \\
          &       &       & \multicolumn{1}{l}{(3.0816)} &       &       &       & \multicolumn{1}{l}{(2.8787)} &  \\
    \multirow{2}[0]{*}{ln(Innovation)} &       &       & \multicolumn{1}{l}{0.0458*} &       &       &       & \multicolumn{1}{l}{0.0306} &  \\
          &       &       & \multicolumn{1}{l}{(0.0185)} &       &       &       & \multicolumn{1}{l}{(0.0192)} &  \\
    \multirow{2}[1]{*}{ln(Trade)} &       &       &       & \multicolumn{1}{l}{0.1107***} &       &       &       & \multicolumn{1}{l}{0.1101***} \\
          &       &       &       & \multicolumn{1}{l}{(0.0180)} &       &       &       & \multicolumn{1}{l}{(0.0175)} \\
    \hline
    Observations & 186   & 186   & 186   & 186   & 186   & 186   & 186   & 186 \\
    Adjusted $R^2$ & 0.4933 & 0.5584 & 0.6277 & 0.5794 & 0.4961 & 0.5302 & 0.6402 & 0.5852 \\
    RMSE  & 0.3174 & 0.2963 & 0.2720 & 0.2891 & 0.3165 & 0.3056 & 0.2674 & 0.2871 \\
    \hline
    \end{tabular*}
    \begin{flushleft}
    \emph{Notes}: These multivariate regressions use the ordinary least squares (OLS) models to regress the level of economic development (GDP pc) against Economic Complexity Index (ECI) in columns (1)-(4) and Fitness Index in columns (5)-(8). These regressions include year-fixed effects using data in the 2010-2015 period. Regression coefficients of variables with standard errors (in the corresponding parentheses) are reported under significant level *$p<0.1$, **$p<0.05$, and ***$p<0.01$. Adjusted $R^2$ indicates how many data points fall within the line of the regression equation, and RMSE stands for the root mean square error.
    \end{flushleft}
  \label{Tab1}
\end{table*}

Table~\ref{Tab1} summarizes the results of multivariate regressions by using ordinary least squares (OLS) models with year-fixed effects for the period 2010-2015. The dependent variable is ln(GDP pc) and the independent variables of interest are ECI for columns (1)-(4) and Fitness for columns (5)-(8) of Table~\ref{Tab1}. We find that ECI is a positive and significant indicator for the level of economic development, and it solely explains 49.33\% of the variance in ln(GDP pc) among provinces (see column (1) of Table~\ref{Tab1}). Also, Urbanization shows positive and significant relationship with ln(GDP pc). The factor ln(Population) has positive correlation with ln(GDP pc), but the result is not significant (see column (2) of Table~\ref{Tab1}). These three factors can explain 55.84\% of the variance in ln(GDP pc).

Economic research has revealed the importance of education, which raises people's knowledge, skills, productivity and creativity, as a crucial and fundamental factor in economic development \cite{Nelson1966,Cainelli2006,Autor2014}. Here, we find that both Schooling (the ratio of students in higher education in a province) and Innovation (the number of domestic granted patents) are positively and significantly correlated with ln(GDP pc), as shown in column (3) of Table~\ref{Tab1}. The explanatory power of ECI remains positive and significant after controlling the effects of Schooling and Innovation. The three factors together explain up to 62.77\% of the variance in ln(GDP pc). In column (4) of Table~\ref{Tab1}, we find the positive and significant correlation between ln(GDP pc) and Trade (the total value of imports and exports of foreign trade). ECI and ln(Trade) can explain 57.94\% of the variance in ln(GDP pc).

Columns (5)-(8) of Table~\ref{Tab1} present regression results using Fitness, where we find Fitness alone has the close explanatory power as ECI (see column (1) and column (5)). However, one should notice that Fitness and ECI are distinguishable from each other, for example, their formulas have essential differences (see Eq.~(\ref{Eq:ECI}) and Eq.~(\ref{Eq:Fit})), the distributions of the values that they produce are different (see bar plots in the upper-left of Figure~\ref{Fig6}), and the correlation between their values is 0.872 instead of 1 (see scatter plot and correlation value in the upper-left of Figure~\ref{Fig6}). Indeed, after controlling the effects of Population and Urbanization, the explanatory power of Fitness is slightly inferior to ECI, as indicated by the smaller values of Adjusted-$R^2$ (see column (6) of Table~\ref{Tab1}). Also, Fitness becomes more powerful than ECI, after controlling the effects of Schooling, Innovation and Trade (see columns (7) and (8) of Table~\ref{Tab1}). Moreover, we notice that ln(Innovation) loses its explanatory power for ln(GDP pc) in the Fitness regression, as shown in column (7) of Table~\ref{Tab1}. In short, ECI and Fitness are comparative with each other, and both of them are robust in explaining regional economic development.

\section{Conclusions and Discussion}
\label{sec4}

In this paper, we studied China's regional economic complexity based on 25 years' firm data covering 31 provinces and 70 industries. First, we mapped the firm data to a ``Province-Industry'' bipartite network, based on which we found that provinces with a high level of economic diversity trend to have the comparative advantages in industries with a low level of ubiquity. Then, we quantified the competitiveness of provinces through the non-monetary Economic Complexity Index (ECI) by defining a set of linear iterative equations between provinces' economic complexity and industries' ubiquity. We found that provinces located around the coast have larger ECIs, and the overall time evolution of provinces' rankings by ECI are relatively stable and slow. Further, after linking ECI with the economic development, as measured by GDP pc and the relative income at urban (RICU) and rural areas (RICR), and the relative income differences (RICD), we found that ECI is positively and significantly correlated with the level of economic development while negatively correlated with income inequality, suggesting that ECI has potential to be a good non-monetary indicator for revealing the status of regional economic development.

Moreover, we compared different measures of non-monetary economic complexity and diversity themselves (ECI, Fitness, Diversity and Entropy), and explored their relationships with some traditional monetary macroeconomic indicators (GDP pc, RICU, RICR and RICD). We found that both ECI and Fitness have higher and positive correlations with the level of economic development, compared to Diversity and Entropy. Together, we found the relative income in rural area (RICR) outperforms the relative income in urban area (RICU) in correlating with the economic diversity measures. Moreover, we showed that all the measures of economic diversity and economic development are negatively and significantly correlated with the income inequality (RICD), suggesting that provinces with higher economic diversity and relative income have less income inequality. Finally, we checked the robustness of the explanatory power of ECI and Fitness for economic development using multivariate regressions with controlling for the effects of some socioeconomic factors like Population, Urbanization, Schooling, Innovation and Trade. Results suggest that both ECI and Fitness are robust in correlating with regional economic development. Even though the causal relation between economic complexity and development cannot be established yet, our work still contributes to the literature on the complexity of regional economic systems within a nation.

Nevertheless, our results are not beyond limitations on data and modeling. The firm data contain a tiny fraction of all Chinese firms. In fact, some very successfully and representative firms are not included in our analysis just because they are not listed in the two major stock markets. Also, our data are limited by the spatial resolution because provinces of China are first-level administrative divisions with heterogeneous land area. Some provinces have large area but small population like Inner Mongolia, while some may be opposite like Jiangsu. Moreover, the ``Province-Industry'' network is built by counting how many firms in one province that operate in an industry without considering the revenues and sizes of firms. This may cause potential biases, to some extent, towards small firms since they have less economic capacity, yet small contribution to regional economy development. In addition, the emergence of new industries is limited by whether they have the comparative advantage, which is not intuitive than by the absolute number of firms. Furthermore, our current analysis is unable to establish causal relation between economic complexity and development, limiting prediction as correlation in this context. Besides, due to the lack of officially reported panel data of Gini coefficients for Chinese provinces, the relative income differences in urban and rural areas in 2010 is used as an alternative in estimating income inequality \cite{Xie2014}, which limits the time evolution analysis and the comparison with other literature that used Gini coefficients. These aforementioned limitations call for improvements towards better understanding the status of regional economic development of China during its period of economic expansion.

How to better quantify economic complexity in both theoretical and empirical ways is still an open question, which remains further investigation. For example, as pointed out by previous studies \cite{Cristelli2013,Cristelli2015,Tacchella2012}, the two main economic complexity indicators sometimes don't show consistency with each other in ranking countries based on world trade data, and traditional regression analysis is not particularly meaningful for addressing the economic complexity problem. However, due to the lack of ground truth and the dependency on dataset in empirical studies, arguments on which indicator performs best and which branch of theories is the most suitable to address this problem will not see their ends and now urge on quantitative evaluation methods \cite{Mariani2015}. Nevertheless, in recent years this branch of economic complexity studies have found widely applications in ranking countries, industries, institutions, occupations and products \cite{Hartmann2017,Hausmann2014,Cristelli2015}, see for example the Observatory of Economic Complexity (OEC) (http://atlas.media.mit.edu), the DataViva (http://www.dataviva.info), and the Growthcom (http://www.growthcom.eu). Meanwhile, the widely application at different scales challenges the practicability of these methods, for example, the consistency or inconsistency of ECI and Fitness results at national level and regional level. Keeping these potential limitations and promising real world applications in mind, we would leave seeking data covering more firms with higher spatial resolution, checking the robustness of findings using alternative definitions of new industry presence, exploring new methods to evaluate the performance of economic complexity indicators and proposing novel economic complexity metrics at different scales as future works.

Indeed, the increasing complexity of economic systems and the data revolution of the past decade urge us on a paradigm change in a more complexity-oriented and data-oriented economic thinking \cite{Martin2007,Durlauf2005,Cristelli2013b}. For example, mainstream approaches measure economic development and predict economic growth using the aggregated GDP based on economic census, financial market, foreign investment, physical capital, and so on \cite{Lucas1988,Mankiw1992,Levine1998}. However, computing monetary factors, for instance GDP, is usually a non-trivial task due to their involvement with considerable resources for a long period \cite{Liu2016}. In recent years, as the availability of large-scale data \cite{Einav2014,Einav2014b} and the development of complexity \cite{Hausmann2014,Hidalgo2015} and network science \cite{Newman2010,Lewis2011,Barabasi2016}, new conceptual frameworks have been developed to address these issues in a more efficient way with far less cost. For example, based on world trade data, ``Product Space'' was proposed which reveals the status of national economic development and explains why not all countries face the same opportunities in future development \cite{Hidalgo2007}, and non-monetary economic complexity and fitness were introduced which have potential to predict future growth \cite{Hidalgo2009}. Moreover, online social networks \cite{Liu2016,Yuan2016,Gao2014}, mobile phone data \cite{Eagle2010,Blumenstock2015,Vscepanovic2015}, satellite imagery \cite{Jean2016,Doll2006}, geo-tagged images \cite{Salesses2013} and web queries \cite{Preis2010,Choi2012} have also been applied to reveal economic status, infer economic development, forecast unemployment, predict poverty, map inequality, quantify trading behavior \cite{Preis2013} and correlate stock market moves \cite{Curme2014}. Although the new way of economic thinking is not perfect \cite{Cristelli2013,Cristelli2015} and somehow limited by the availability of data and new statistical tools, there is a high possibility that it will change the landscape of economic research in the near future \cite{Einav2014b}.

\vspace{6pt}

\section*{Acknowledgments}
The authors acknowledge the anonymous reviewers for critical comments and constructive suggestions. The authors thank Haixing Dai, Yiding Liu, Zhihai Rong, Qing Wang, and Dan Yang for helpful discussions. This work was partially supported by the National Natural Science Foundation of China (Grant Nos. 61433014 and 61673086). Jian Gao acknowledges the China Scholarship Council for partial financial support and the Collective Learning group at the MIT Media Lab for hosting.

\appendix{}
\setcounter{table}{0}
\renewcommand{\thetable}{A\arabic{table}}
\section{}

\begin{table}[!ht]
  \centering
  \caption{The two-digital abbreviations of province names in China.}
  \footnotesize
    \begin{tabular*}{\textwidth}{@{\extracolsep{\fill}}ccl|ccl|ccl}
    \toprule
    ID    & Abbreviation & \multicolumn{1}{c|}{Province} & ID    & Abbreviation & \multicolumn{1}{c|}{Province} & ID    & Abbreviation & \multicolumn{1}{c}{Province} \\
    \midrule
    1     & BJ    & Beijing & 12    & AH    & Anhui & 23    & SC    & Sichuan \\
    2     & TJ    & Tianjin & 13    & FJ    & Fujian & 24    & GZ    & Guizhou \\
    3     & HE    & Hebei & 14    & JX    & Jiangxi & 25    & YN    & Yunnan \\
    4     & SX    & Shanxi & 15    & SD    & Shandong & 26    & XZ    & Tibet \\
    5     & NM    & Inner Mongolia & 16    & HA    & Henan & 27    & SN    & Shaanxi \\
    6     & LN    & Liaoning & 17    & HB    & Hubei & 28    & GS    & Gansu \\
    7     & JL    & Jilin & 18    & HN    & Hunan & 29    & QH    & Qinghai \\
    8     & HL    & Heilongjiang & 19    & GD    & Guangdong & 30    & NX    & Ningxia \\
    9     & SH    & Shanghai & 20    & GX    & Guangxi & 31    & XJ    & Xinjiang \\
    10    & JS    & Jiangsu & 21    & HI    & Hainan &       &       &  \\
    11    & ZJ    & Zhejiang & 22    & CQ    & Chongqing &       &       &  \\
    \bottomrule
    \end{tabular*}
  \label{Tab:A1}
\end{table}


\end{document}